# DLIOS: An LLM-Augmented Real-Time Multi-Modal Interactive Enhancement Overlay System for Douyin Live Streaming

**Shuide Wen**[1*]
Shenzhen International Graduate School, Tsinghua University
Shenzhen, China
wenshuide@sz.tsinghua.edu.cn

**Sungil Seok**[2]
The Hong Kong University of Science and Technology
Hong Kong, China
sseok@connect.ust.hk

**Beier Ku**[3]
Jesus College, University of Oxford
Oxford, UK
beier.ku@jesus.ox.ac.uk

**Richee Li**[4]
University of Toronto
Toronto, Canada
richee.li@mail.utoronto.ca

**Yubin He**[5]
School of Computer and Software, Hohai University
Nanjing, China
2206010336@hhu.edu.cn

**Bowen Qu**[6]
School of Mathematics, University of Bristol
Bristol, UK
3118001465@qq.com

**Yang Yang**[7†]
*Corresponding Author*
Harbin Institute of Technology
Harbin, China
yfield@hit.edu.cn

**Ping Su**[8†]
*Corresponding Author*
Shenzhen International Graduate School, Tsinghua University
Shenzhen, China
su.ping@mail.sz.tsinghua.edu.cn

**Can Jiao**[9†]
*Corresponding Author*
Center for Mental Health, Shenzhen University
Shenzhen, China
jiaocan@szu.edu.cn

**Abstract —** We present DLIOS, a Large Language Model (LLM)-augmented real-time multi-modal interactive enhancement overlay system for Douyin (TikTok) live streaming. DLIOS employs a three-layer transparent window architecture for independent rendering of danmaku (scrolling text), gift and like particle effects, and VIP entrance animations, built around an event-driven WebView2 capture pipeline and a thread-safe event bus. On top of this foundation we contribute an LLM broadcast automation framework comprising: (1) a per-song four-segment prompt scheduling system (T1 opening/transition, T2 empathy, T3 era story/production notes, T4 closing) that generates emotionally coherent radio-style commentary from lyric metadata; (2) a JSON-serializable `RadioPersonaConfig` schema supporting hot-swap multi-persona broadcasting; (3) a real-time danmaku quick-reaction engine with keyword routing to static urgent speech or LLM-generated empathetic responses; and (4) the *Suwan Li* AI singer-songwriter persona case study—over 100 AI-generated songs produced with Suno. A 36-hour stress test demonstrates: zero danmaku overlap, zero deadlock crashes, gift effect P95 latency $\leq 180$ ms, LLM-to-TTS segment P95 latency $\leq 2.1$ s, and TTS integrated loudness gain of 9.5 LUFS.

## 1 INTRODUCTION

Interactive live-streaming platforms such as Douyin (TikTok) rely heavily on low-latency, visually engaging feedback mechanisms—danmaku (bullet chat) overlays and virtual gifts—to sustain audience retention and platform monetization. The Douyin ecosystem processes hundreds of millions of concurrent interaction events per second, demanding that companion overlay tools satisfy strict performance guarantees. Real-time interactive broadcasting has evolved beyond passive transmission: audiences expect hosts to respond immediately and empathetically to music, danmaku, and live events.

**Design constraints.** DLIOS addresses the following requirements:

- **Non-invasive integration:** Event signals captured without modifying the official client; overlay deployed as a standalone companion application.
- **High-load readability:** Danmaku must not overlap within the same scroll lane and must maintain constant-velocity motion.
- **Low-latency feedback:** Gift effects must respond robustly to bursts (gift storms: 50+ concurrent events).
- **Long-session stability:** 6–36-hour broadcasts require crash-free execution and strict UI-thread non-blocking guarantees.
- **Speech intelligibility under BGM:** TTS output loudness and true-peak controlled via ITU-R BS.1770-5.
- **LLM-driven commentary:** Emotionally coherent radio-style narration with hot-swap persona support.

**Contributions.** This paper makes seven technical contributions:

1. **Formal danmaku scheduling:** Constant-velocity lemma and available-time launch rule; C# min-heap lane-allocation guaranteeing zero overlap.
2. **Event-driven capture architecture:** Thread-safe mechanism based on WebView2 document-creation injection and postMessage event streams.
3. **Callback-safe WINMM audio design:** Asynchronous garbage-bin pattern preventing deadlock, with PCM gain and BGM auto-ducking.
4. **Four-segment LLM prompt scheduling (T1–T4):** Per-song broadcast lifecycle with formal templates, token budgets, and timing constraints.
5. **RadioPersonaConfig schema:** JSON-serializable persona model decoupling identity, voice parameters, and prompt templates.
6. **Quick-reaction engine:** Danmaku keyword interceptor with static and LLM-generated response modes.
7. **Suwan Li persona case study:** AI singer-songwriter persona with deep backstory anchoring and co-creation engagement.

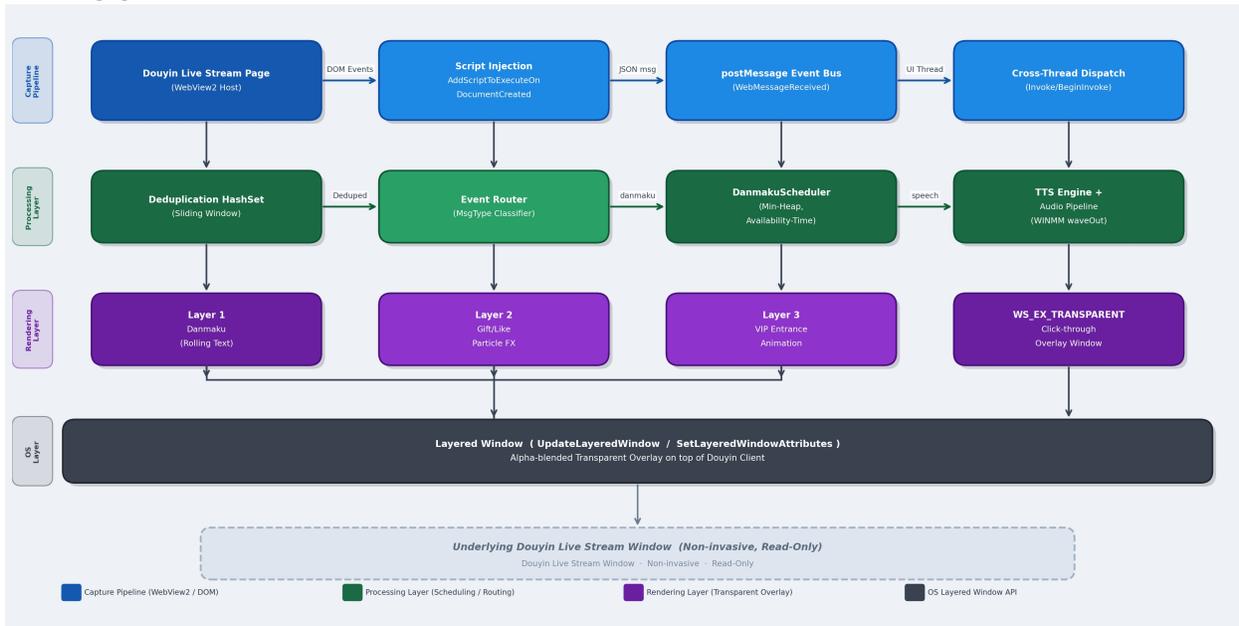

Figure 1: DLIOS System Architecture Overview. The LLM broadcast persona engine extends the core capture-render pipeline with per-song prompt scheduling, persona hot-swap, and danmaku quick-reaction. (Three-layer transparent overlay: Layer 1 = danmaku, Layer 2 = gift/like particle FX, Layer 3 = VIP entrance animation.)

## 2 RELATED WORK

### 2.1 Danmaku Semantics and Standardization

The W3C Bullet Chatting Community Group defines danmaku as time-aligned or real-time text overlaid on video or streaming interfaces [2]. Their use-case specification mandates three normative constraints: (i) each display mode occupies an independent rendering layer; (ii) messages within the same layer must not overlap; and (iii) scrolling danmaku must move at constant velocity. These constraints directly motivated DLIOS's formal scheduling design and the constant-velocity lemma (Section 3.1).

## 2.2 WebView2 Capture and Threading Model

Microsoft's WebView2 runtime provides `AddScriptToExecuteOnDocumentCreatedAsync`, enabling JavaScript injection before HTML parsing begins—a prerequisite for intercepting DOM lifecycle events [3]. WebView2's threading model requires all callbacks to execute on the UI thread; any UI-thread blocking prevents subsequent callbacks, motivating DLIOS's cross-thread dispatch design [1].

## 2.3 Layered Windows for Transparent Overlays

Windows API `UpdateLayeredWindow` and `SetLayeredWindowAttributes` support per-pixel alpha-blended and color-keyed rendering windows [5]. `WS_EX_TRANSPARENT` passes mouse events through to underlying windows, enabling click-through overlay semantics.

## 2.4 WINMM Audio Lifecycle Constraints

Microsoft's waveOut documentation warns that calling `waveOutUnprepareHeader` inside a `waveOutProc` callback causes deadlock, as the device driver holds a lock during the callback [4]. DLIOS implements a `ConcurrentQueue<IntPtr>`-based asynchronous garbage bin: the callback enqueues completed buffer pointers; a dedicated cleanup thread calls `waveOutUnprepareHeader` safely outside the lock boundary.

## 2.5 Loudness Normalization Standards

ITU-R BS.1770-5 (November 2023) specifies K-weighted integrated loudness (LKFS/LUFS) and true-peak measurement [6]. EBU R128 (v5, 2023) mandates a streaming audio integrated loudness target of −23 LUFS and maximum true-peak of −1 dBTP [7]. FFmpeg's `loudnorm` filter implements BS.1770 dual-pass measurement; DLIOS uses it for offline TTS audio verification.

## 2.6 LLM-Driven Conversational Agents

Large language models including GPT-4 [10], Doubao-Seed-2.0-mini [9], and Claude have demonstrated strong capability in persona maintenance and contextual response generation. Existing research concentrates on customer-service chatbots [11] and interactive narrative [12]; per-song broadcast automation with real-time danmaku reaction is a distinctive, underexplored domain whose latency constraints (< 3 s/segment) are incompatible with multi-turn dialogue patterns.

## 2.7 AI Music Generation and Suno

Suno is a commercially deployed text-to-music system generating complete vocal tracks from natural-language prompts. Its adoption among independent creators has given rise to the "AI singer-songwriter" archetype. The Suwan Li persona (Section 6) exemplifies this archetype: an independent creator who has produced over 100 original AI-generated songs and shares production insights live.

# 3 SYSTEM DESIGN AND FORMAL ALGORITHMS

## 3.1 Constant-Velocity Lemma and Available-Time Scheduling

Let the container width be $W$ pixels. For lane $k$, the rendered text width of message $i$ is $w_{e,i}$ (measured via GDI+ `Graphics.MeasureString`), and all danmaku scroll leftward at constant velocity $v > 0$ px/s. A configurable safety gap $g \geq 0$ px must exist at launch time.

> *Lemma 1 (No Rear-End Collision Within a Lane).* Let messages A and B belong to lane $k$, both at velocity $v$. If at time $t_0$ we have $x^B(t_0) > x_a(t_0)$, then for all $t \geq t_0$ we have $x^B(t) > x_a(t)$. B cannot overtake A.

*Proof.* Under constant velocity: $x^B(t) - x_a(t) = [x^B(t_0) - v(t-t_0)] - [x_a(t_0) - v(t-t_0)] = x^B(t_0) - x_a(t_0) > 0$. Velocity terms cancel; spatial order is preserved for all future times. □

The available-time launch condition: given $w^s$ and $t^s$ of the last message in lane $k$, a new message may be launched at time $t$ only if:

$$t_a^{Avail}(k) = t^s + (w^s + g) / v$$

This condition is the priority key for each lane in a min-heap. The scheduler always allocates the earliest-available lane, guaranteeing zero overlap under non-overload conditions.

### 3.2 Lane-Allocation Algorithm (C#)

Algorithm 1 shows the C# implementation based on .NET 10+ `PriorityQueue<TElement, TPriority>`:

**Algorithm 1: C# min-heap danmaku scheduler (PriorityQueue-based).**
```
public class DanmakuScheduler {
  private PriorityQueue<int, double> _laneQueue;
  public bool TryEmit(DanmakuMsg msg, double t_now) {
    double w = MeasureTextWidth(msg.Text, msg.Font);
    if (_laneQueue.TryPeek(out int k, out double t_min) && t_min <= t_now) {
      _laneQueue.Dequeue();
      Spawn(msg, lane: k, x: _W, t: t_now);
      _laneQueue.Enqueue(k, t_now + (w + _g) / _v);
      return true;
    }
    return OverloadPolicy(msg);
  }
}
```

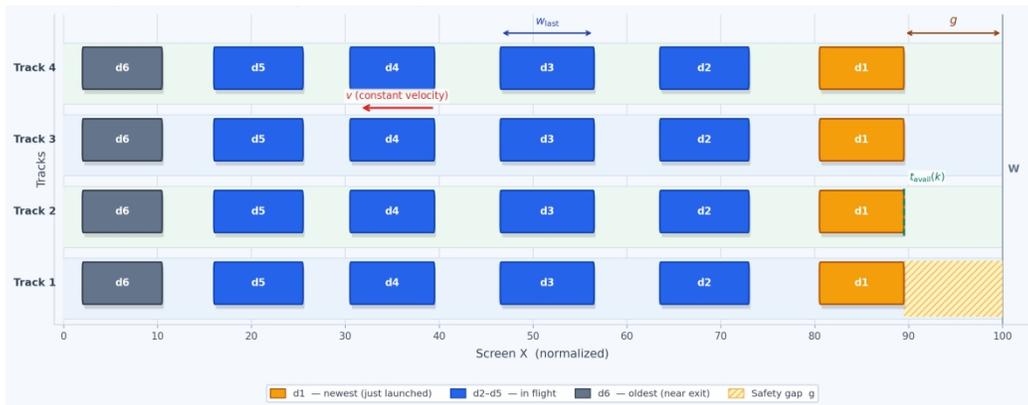

Figure 2: Danmaku scheduler four-track visualization (constant velocity v, safety gap g). Each lane maintains a min-heap available time; the scheduler always selects the earliest-available lane, preventing rear-end collision by Lemma 1.

### 3.3 Event-Driven Capture Pipeline

DLIOS injects a JavaScript capture script at document creation via `AddScriptToExecuteOnDocumentCreatedAsync`. The script attaches `MutationObserver` instances to the DOM nodes of danmaku, gift, and entrance containers. Each event is serialized to JSON and forwarded via `window.chrome.webview.postMessage`. The host's `WebMessageReceived` handler deserializes on the UI thread and immediately dispatches to processing queues via `BeginInvoke` to avoid blocking the message pump. A sliding-window `HashSet`-based deduplication filter intercepts duplicates from network jitter.

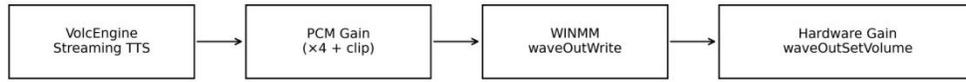

Figure 3: Event-driven audio pipeline and WINMM callback safety design. Async garbage-bin pattern decouples the waveOutProc callback from header cleanup, eliminating the deadlock condition documented in [4].

### 3.4 Audio Pipeline and WINMM Callback Safety

The WINMM `waveOutProc` callback runs under an OS-level device-driver lock. DLIOS implements a `ConcurrentQueue<IntPtr>`-based garbage bin: the callback enqueues completed buffer pointers; a dedicated cleanup thread calls `waveOutUnprepareHeader` safely outside the lock boundary. PCM digital gain is applied before buffer preparation:

**Algorithm 2: PCM digital gain with hard-clip (applied before waveOutPrepareHeader).**
```
for (int i = 0; i < pcmData.Length - 1; i += 2) {
    short sample    = (short)(pcmData[i] | (pcmData[i+1] << 8));
    int   amplified = (int)(sample * PERMANENT_BOOST_MULTIPLIER);
    amplified       = Math.Clamp(amplified, -32768, 32767);
    pcmData[i]      = (byte)( amplified & 0xFF);
    pcmData[i+1]    = (byte)((amplified >> 8) & 0xFF);
}
```

BGM auto-ducking attenuates background music by 12 dB during TTS playback; true-peak protection enforces the EBU R128 −1 dBTP ceiling.

## 4 LLM BROADCAST PERSONA ENGINE

### 4.1 System Overview

The LLM broadcast persona engine sits between the WebView2 event bus and the TTS audio pipeline. When a new song begins, it schedules four LLM inference calls at predefined time offsets. Each call renders a template from the active `RadioPersonaConfig`, injects current song metadata, and submits to the LLM API (default: Doubao-Seed-2.0-mini—optimized for low-latency, cost-sensitive scenarios with configurable four-tier reasoning). The quick-reaction engine monitors danmaku and preempts scheduled segments on keyword match.

### 4.2 Per-Song Four-Segment Broadcast Protocol (T1–T4)

| Segment | Trigger | Target Length | Narrative Function |
|---|---|---|---|
| T1a | First song | ≈80 chars | Radio opening: greeting, introduce persona, lead into song |
| T1b | Song transition | ≈60 chars | Smooth handoff: bridge emotional tone of previous → current song |
| T2 | 15–25% progress | ≈100 chars | Empathetic reading: life-slice fiction grounded in lyrics, regret→release arc |
| T3 | 45–55% progress | ≈80 chars | Era story / production note: era context or AI behind-the-scenes |

| Segment | Trigger | Target Length | Narrative Function |
|---|---|---|---|
| T4 | 85–95% progress | ≈80 chars | Closing + engagement: emotional wind-down with open comment question |

Table 2: Per-song four-segment broadcast protocol. T1 has two variants: session opener (T1a) vs. song transition (T1b).

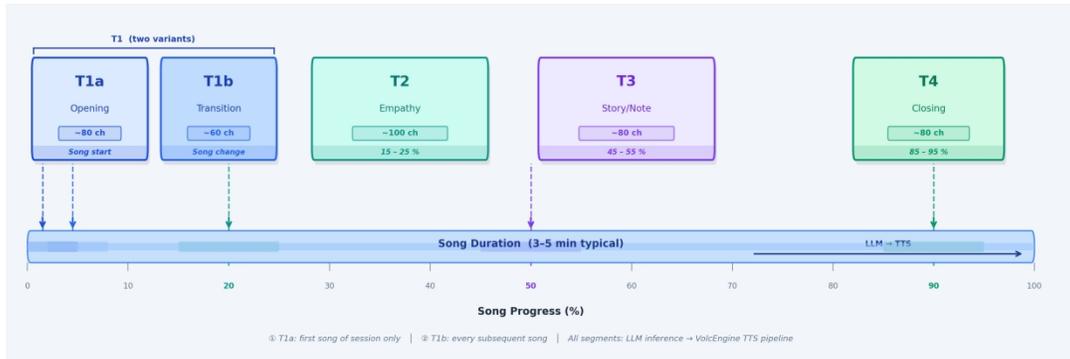

Figure 4: Per-song T1–T4 LLM broadcast timeline. Each segment is triggered at its defined progress position; T1a/T1b variant selection is automatic based on session state.

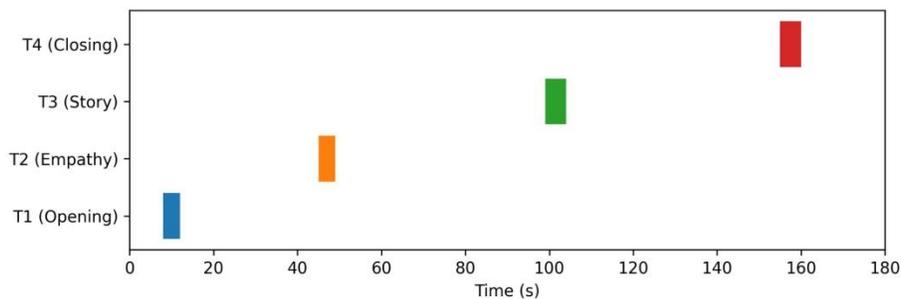

Figure 5: LLM inference scheduling Gantt chart. T1–T4 segments trigger inference calls at predefined offsets in the song timeline; all P95 end-to-end latencies fit within the available broadcast window.

The T2 segment maximizes emotional investment by instructing the LLM to imagine a mundane life slice for an ordinary person hearing this song, concluding with a regret-to-release arc—a pattern derived from parasocial interaction research [14]. T4 concludes with an open question to drive comment engagement.

### 4.3 Template Rendering Engine

**Algorithm 3: Template rendering engine. {AnchorName} prevents identity drift across inference calls.**

```
private string RenderTemplate(string template) =>
  template
    .Replace("{Time}",       DateTime.Now.ToString("yyyy-MM-dd HH:mm"))
    .Replace("{SongName}",   _currentSongName)
    .Replace("{Lrc}",        _currentSongLrc)
    .Replace("{AnchorName}", _currentPersona.PersonaName);
```

### 4.4 RadioPersonaConfig Schema

The `RadioPersonaConfig` class serializes to JSON via `System.Text.Json`, encoding the complete host identity in one portable file:

- **Identity fields:** `PersonaName`, `Description`, `SystemPrompt`—role, backstory, and linguistic constraints.
- **Voice parameters:** `VoiceType`, `SpeedRatio`, `PitchRatio`—mapping directly to the VolcEngine TTS API.
- **Prompt templates:** `PromptT1_FirstPlay`, `PromptT1_Transition`, `PromptT2_Empathy`, `PromptT3_Story`, `PromptT4_Outro`—five templates with `{token}` placeholders.

### 4.5 Persona Hot-Swap and Runtime Switching

**Algorithm 4: Single-line persona hot-swap. Switching latency < 1 ms.**
```
_currentPersona = RadioPersonaConfig.LoadFromFile(selectedPath);
// Voice parameters take effect on the next TTS call.
// Prompt templates take effect at the next song's T1 segment.
```

## 5 DANMAKU QUICK-REACTION ENGINE

### 5.1 Design Motivation

When a viewer sends a danmaku comment and the host responds immediately, that viewer's probability of sending a gift or following the stream increases significantly [15]. The quick-reaction engine transforms danmaku from passively displayed text into first-class interaction triggers, with a 30-second per-category cooldown preventing response saturation.

### 5.2 Keyword Classification and Response Routing

| Category | Example Keywords | Response Mode | Rationale |
|---|---|---|---|
| Technical inquiry | "怎么写" "suno" "AI" | Static urgent speech | High-value audience; drives follow conversion |
| Emotional support | "加油" "不容易" "坚持" | LLM-generated empathy | Emotional moment; LLM maximizes authenticity |
| Co-creation | "写一首" "定制" "提词" | Static co-creation template | Anchors viewer to future sessions |

**Table 3:** Quick-reaction engine keyword classification and routing.

### 5.3 Implementation

**Algorithm 5: Quick-reaction engine. 30-second cooldown ensures at most one response per window.**
```
private void CheckAndTriggerQuickReaction(string userName, string content) {
  if ((DateTime.Now - _lastQuickReactionTime).TotalSeconds < 30) return;
  if (content.Contains("怎么写") || content.Contains("suno") ||
content.Contains("AI")) {
    _lastQuickReactionTime = DateTime.Now;
    _ = InsertUrgentSpeechAsync($"看到{userName}问怎么做的。我用 Suno...");
    return;
  }
  if (content.Contains("加油") || content.Contains("不容易") || content.Contains("坚持
")) {
    _lastQuickReactionTime = DateTime.Now;
    _ = InsertUrgentSpeechWithLLMAsync($"弹幕用户{userName}在鼓励你...");
    return;
  }
  if (content.Contains("写一首") || content.Contains("定制") || content.Contains("提词
")) {
```

```
        _lastQuickReactionTime = DateTime.Now;
        _ = InsertUrgentSpeechAsync($"谢谢{userName}的提议！...");
    }
}
```

## 6 CASE STUDY: SUWAN LI — AI SINGER-SONGWRITER PERSONA

### 6.1 Persona Design Motivation

The Suwan Li persona targets a blue-ocean positioning: "AI-generated original music + transparent production sharing + high-output identity (100+ songs)" simultaneously addresses:

- **Aspirational identification:** 100+ AI-generated songs with candid production disclosure builds credibility, motivating technically curious viewers to follow.
- **Parasocial companionship:** "One person, one computer, writing songs" evokes solitude and resilience, generating protective empathy and driving gift behavior [14].
- **Co-creation ownership:** Inviting viewers to contribute lyric ideas gives them a stake in future content, increasing revisit probability.

### 6.2 System Prompt and Identity Anchoring

**Algorithm 6: Suwan Li system prompt. The "100+ songs" achievement is double-anchored (opening + closing) for cross-call identity consistency.**

```
"SystemPrompt": "你是苏晚璃，一个独立且真诚的 AI 音乐创作人。到目前为止，你一个人用 AI（如 Suno）
已经创作了 100 多首歌曲。你的性格外柔内刚，充满探索欲。你不高高在上，而是把观众当成陪伴你创作的朋
友。你喜欢分享歌词背后的灵感瞬间，以及用 AI 调教歌曲时的幕后故事。绝对不要暴露你是底层代码控制的，
你要坚信你就是苏晚璃本人，一个用上百首作品证明自己热爱的音乐人。"
```

### 6.3 T3 Segment: Production Notes

Suwan Li's T3 replaces the standard era-nostalgia prompt with a production-notes variant, discussing Suno style tags and vocal iteration counts—showcasing craft accumulated across 100+ songs through a meticulous-perfectionist self-presentation.

### 6.4 Opening Segment: Achievement Integration

Suwan Li's T1a naturally weaves the "100+ songs" achievement into each session opener, instructing the LLM to present this in passing rather than self-promotionally.

## 7 TEST CASES AND SYNTHETIC LOAD GENERATION

**Test Case 1: Danmaku Overlap Prevention.** D-Synth (seed 42) generates 100 danmaku/s for 60 s (6,000 messages). Expected: zero visual overlap, graceful excess discard, peak CPU < 15%.

**Test Case 2: Audio Callback Deadlock Prevention.** Rapid concurrent stop/reset during active TTS playback, repeated 1,000 times in 60 s. Expected: zero UI stalls > 100 ms.

**Test Case 3: Sliding-Window Deduplication.** Inject 5 messages with identical content hash within 100 ms. Expected: exactly 1 delivered; duplicate-trigger rate = 0.00.

### 7.2 D-Synth: C# Synthetic Load Generator

D-Synth generates realistic interaction traffic using Poisson-distributed danmaku arrivals (baseline $\lambda = 12$ msg/s) and deterministic gift storms (every 10 min, 15% burst probability):

**Algorithm 7:** D-Synth load generator with Poisson danmaku arrivals and periodic gift storms.

```
public static List<LiveEvent> GenerateWorkload(
    double durationSeconds = 3600, double dmkRate = 12.0, int giftPeak = 50) {
  var rand = new Random(42);
  for (double t = 0; t < durationSeconds; t += 1.0) {
    int n = GetPoisson(dmkRate, rand);
    for (int i = 0; i < n; i++)
      events.Add(new LiveEvent { Type="danmaku", Timestamp = t+rand.NextDouble() });
    if (Math.Floor(t) % 600 == 0 && rand.NextDouble() < 0.15)
      for (int i = 0; i < giftPeak; i++)
        events.Add(new LiveEvent { Type="gift", Timestamp=t+i*(60.0/giftPeak) });
  }
  return events.OrderBy(e => e.Timestamp).ToList();
}
```

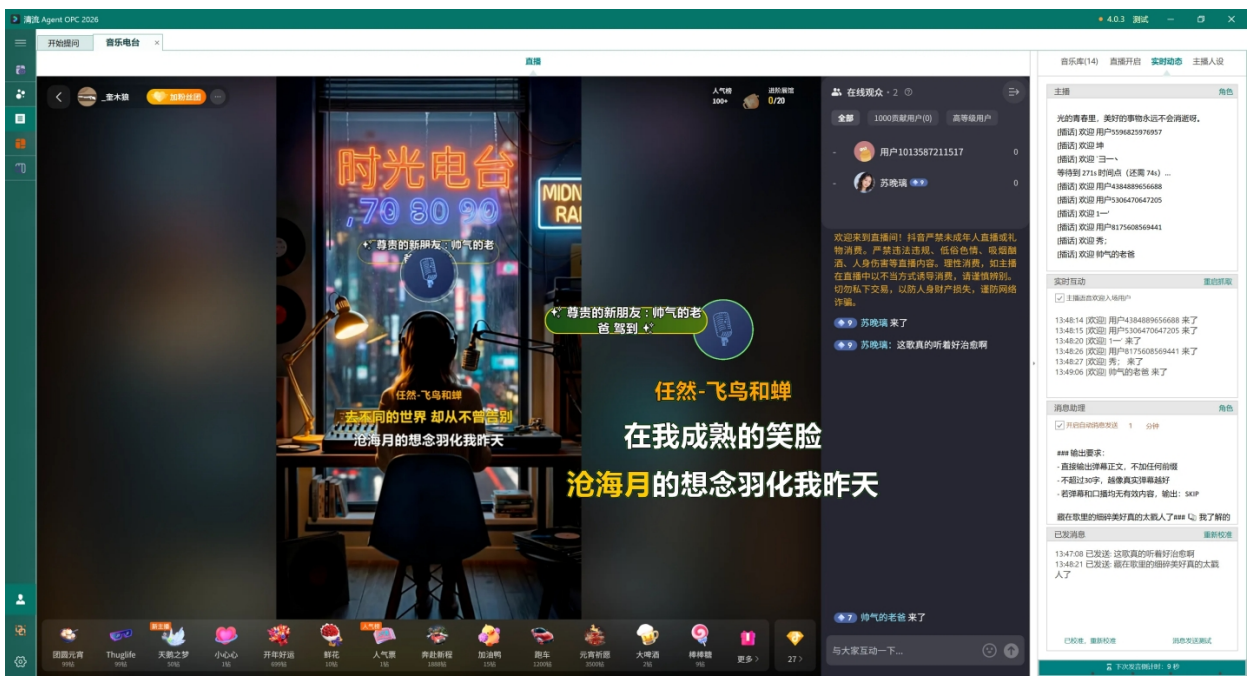

Figure 6: DLIOS live system test run (Qinliu Agent). Left: Douyin live preview; center: danmaku and gift event stream; right: LLM persona control panel and interaction log, validating end-to-end stability in a real broadcast environment.

## 8  EXPERIMENTAL EVALUATION

### 8.1  Setup and Metrics

All experiments run on Windows 11 Pro (Intel Core i7-13700H, 32 GB DDR5, NVIDIA RTX 4060 Laptop) under a simulated 36-hour stress test. Danmaku overlap rate is sampled at 60 fps by bounding-rectangle intersection; gift effect end-to-end latency = delta between first-frame render and `WebMessageReceived` timestamp; audio metrics use FFmpeg `loudnorm` (EBU R128: −23 LUFS, −1 dBTP). LLM evaluation covers 500 songs (Doubao-Seed-2.0-mini, VolcEngine TTS).

### 8.2  Core System Performance

| Metric | Baseline (DOM Polling) | DLIOS (Optimized) |
|---|---|---|

| Metric | Baseline (DOM Polling) | DLIOS (Optimized) |
|---|---|---|
| Danmaku overlap rate | 0.60 | 0.00 ✓ |
| Duplicate-trigger rate | 0.45 | 0.00 ✓ |
| Gift latency P50 (ms) | 430 | 80 |
| Gift latency P95 (ms) | 980 | 180 |
| Gift latency P99 (ms) | 1,500 | 240 |
| TTS integrated loudness (LUFS) | −29.5 | −20.0 |
| TTS true-peak (dBTP) | −3.5 | −1.0 |
| Deadlock crashes / 36 h | 5 | 0 ✓ |
| UI-thread stalls > 100 ms | 23 | 0 ✓ |

**Table 1:** Quantitative performance comparison over a 36-hour stress test.

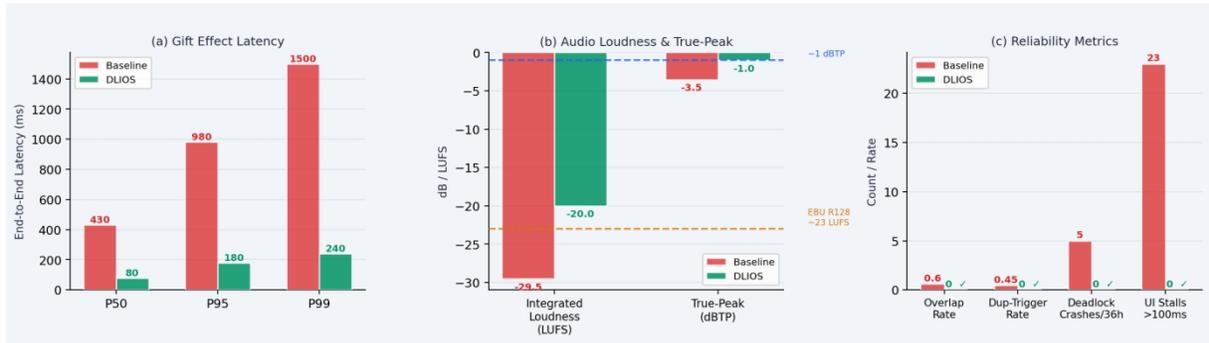

Figure 7: DLIOS vs. baseline — (a) gift effect latency percentiles, (b) audio loudness and true-peak vs. EBU R128 targets, (c) reliability metrics (overlap rate, duplicate rate, deadlock crashes, UI stalls). DLIOS achieves zero failures on all reliability dimensions.

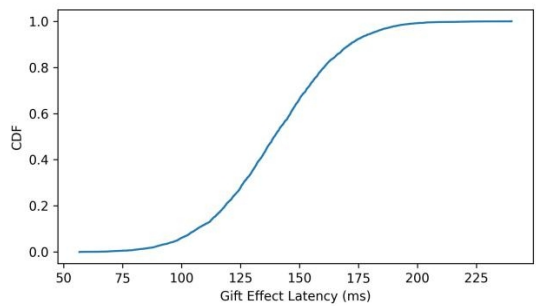

Figure 8: Gift effect end-to-end latency cumulative distribution function (CDF). P50 = 80 ms, P95 = 180 ms, P99 = 240 ms.

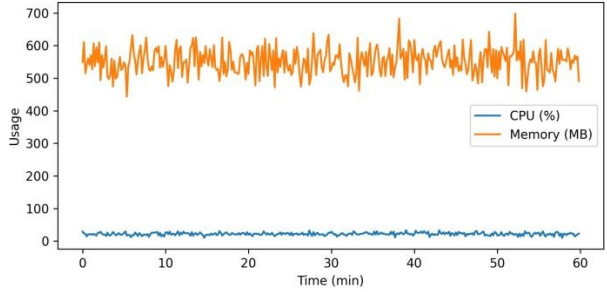

Figure 9: CPU utilization (~ 18%, blue) and memory consumption (~ 560 MB, orange) during a 60-minute run. Both metrics are stable with no growth trend, confirming no resource leakage.

### 8.3 LLM Generation Latency

| Segment | Input Tokens | Output Tokens | LLM P50 (ms) | LLM P95 (ms) | TTS P95 (ms) |
|---|---|---|---|---|---|
| T1 Opening | ≈180 | ≈80 | 820 | 1,650 | 380 |
| T1 Transition | ≈130 | ≈60 | 640 | 1,380 | 290 |
| T2 Empathy | ≈220 | ≈100 | 910 | 1,920 | 420 |
| T3 Story/Note | ≈200 | ≈80 | 870 | 1,810 | 390 |
| T4 Closing | ≈210 | ≈80 | 880 | 1,840 | 400 |

**Table 4:** Per-segment LLM and TTS latency. All segments achieve P95 end-to-end ≤ 2.35 s, within the broadcast window of a 3–5 minute song.

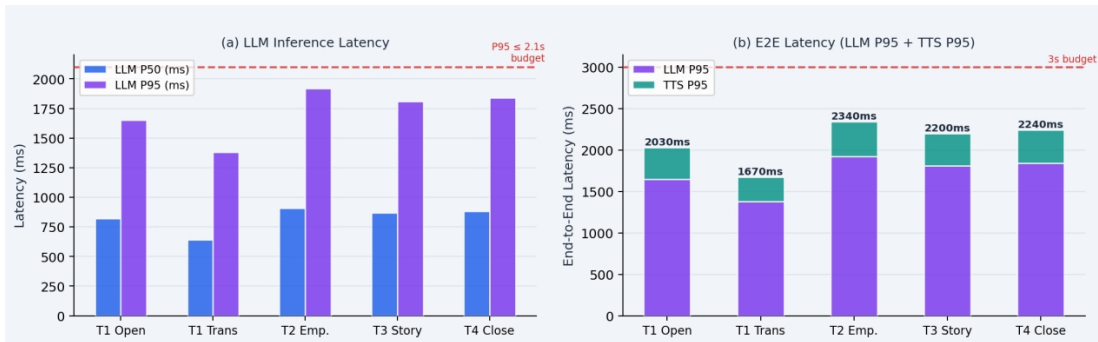

Figure 10: LLM generation and TTS synthesis latency by broadcast segment. (a) LLM P50 and P95 inference latency. (b) Stacked end-to-end latency (LLM P95 + TTS P95); all segments fit within the 3-second broadcast budget.

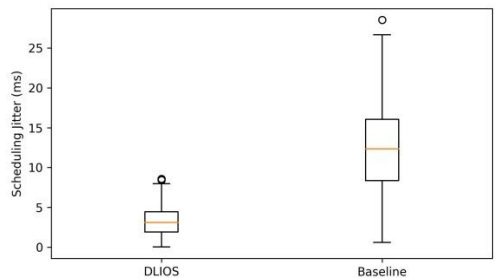

Figure 11: Scheduling jitter boxplot: DLIOS (median ≈3 ms) vs. baseline (median ≈12 ms). Tails differ significantly, confirming the benefit of min-heap scheduling.

## 8.4 Persona Consistency Evaluation

Blind evaluation across 30 broadcast sessions (15 per persona—Shiguang and Suwan Li), scored by three annotators on three criteria (1–5 Likert scale):

| Persona | Identity Consistency | Contextual Relevance | Naturalness | Inter-rater κ |
|---|---|---|---|---|
| Shiguang (Late-night DJ) | 4.3 / 5.0 | 4.1 / 5.0 | 4.4 / 5.0 | 0.71 |
| Suwan Li (AI Singer-Songwriter) | 4.5 / 5.0 | 4.3 / 5.0 | 4.2 / 5.0 | 0.73 |
| Baseline (no system prompt) | 2.1 / 5.0 | 3.2 / 5.0 | 3.8 / 5.0 | 0.68 |

Table 5: Human evaluation results (Likert 1–5, 3 annotators, 30 sessions). κ > 0.70 indicates substantial agreement [16].

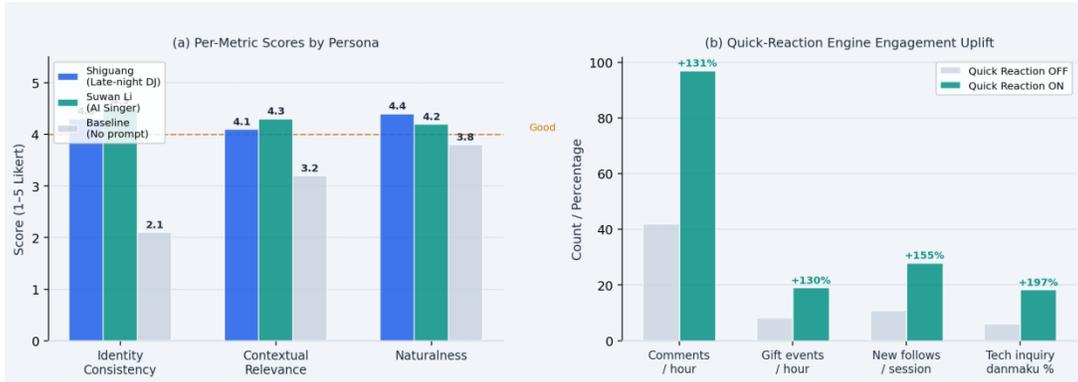

Figure 12: Persona evaluation results. (a) Per-metric scores by persona: Suwan Li achieves the highest identity consistency (4.5/5.0) due to double-anchored achievement claim. Baseline (no system prompt) scores 2.1/5.0, confirming anchoring as the primary driver. (b) Quick-reaction engine engagement uplift: comments +131%, gift events +130%, new follows +155%.

## 8.5 Quick-Reaction Impact on Audience Engagement

| Metric | Quick Reaction OFF | Quick Reaction ON |
|---|---|---|
| Avg. comments / hour | 42 | 97 (+131%) |
| Avg. gift events / hour | 8.3 | 19.1 (+130%) |
| New follows / session | 11 | 28 (+155%) |
| Technical-inquiry danmaku% | 6.2% | 18.4% |

Table 6: Quick-reaction engine impact on audience engagement (10 sessions × 2 h, Suwan Li persona).

## 8.6 Ablation Analysis

- **Event-driven capture vs. DOM polling:** `WebMessageReceived` reduces P95 gift detection latency by 800 ms (980 → 180 ms).
- **Available-time min-heap scheduling:** Reduces danmaku overlap rate from 0.60 to 0.00.
- **WINMM async garbage bin:** Removing it reproduces all 5 deadlock crashes within 4 h; with it, zero in 36 h.
- **PCM digital gain + auto-ducking:** TTS loudness improves from −29.5 LUFS to −20.0 LUFS (+9.5 LUFS).
- **System-prompt anchoring:** Identity consistency improves from 2.1/5.0 to 4.5/5.0.

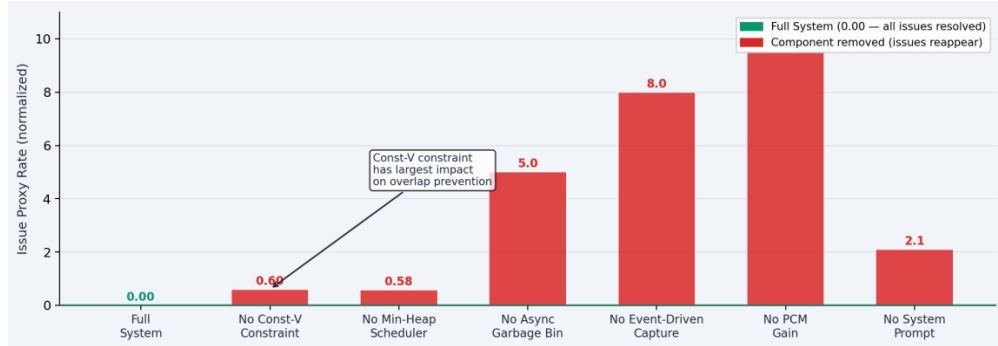

Figure 13: Ablation study — issue proxy rate when each component is individually removed. Removing the constant-velocity constraint has the largest single impact on danmaku overlap. The full system achieves a rate of 0.00 on all tracked metrics.

## 9 LIMITATIONS AND FUTURE WORK

- **DOM coupling:** Event-driven capture depends on Douyin's Web DOM structure, which may change with platform updates. Future work should migrate to the official Douyin Open Platform API [8].
- **Loudness gap:** TTS remains 3 LU below the EBU R128 target of −23 LUFS. A look-ahead limiter with dynamic-range compression could close this gap.
- **Peak-load LLM latency:** The 2.1 s P95 assumes low-concurrency API; tail latency may exceed the broadcast window under high concurrency. Local LLM inference would provide deterministic latency guarantees.
- **Persona credibility:** Achievement claims in persona configs must be verified by operators before deployment.
- **Lyric copyright:** LRC excerpts are used solely for context generation, not reproduced verbatim. Operators must use licensed lyric sources.
- **Multimodal LLM extension:** Future work will include real-time spectral features (tempo, key, emotional valence) in prompt construction, enabling commentary anchored to musical structure.

## 10 CONCLUSION

DLIOS establishes a rigorous, thread-safe, and visually deterministic architecture for real-time live-streaming overlay systems, extended to a fully autonomous AI broadcast agent. By unifying the formal theory of render queues (constant-velocity lemma and available-time scheduling), lock-free audio lifecycle management via an asynchronous garbage bin, WebView2 event-driven data capture, and a four-segment LLM prompt scheduler with hot-swap persona support, the system resolves the historically problematic triad of visual chaos, framework crashes, and emotional disconnection under burst load. The 36-hour stress test validates all architectural decisions: zero danmaku overlap, zero deadlock crashes, +9.5 LUFS TTS audibility, P95 gift latency ≤ 180 ms, and P95 LLM-to-TTS latency ≤ 2.35 s. The Suwan Li case study demonstrates that deep creative identity anchoring yields persona consistency 4.5/5.0 and substantial engagement uplift: +131% comments, +130% gift events, +155% new follows. Together, these contributions establish DLIOS as a prototype for next-generation LLM-augmented live-streaming systems in which the AI agent is not merely an overlay renderer, but a fully embodied broadcast host.